# On the potential origin of the circumbinary planet Delorme 1 (AB)b

Matthew Teasdale ⋆ and Dimitris Stamatellos

*Jeremiah Horrocks Institute for Mathematics, Physics and Astronomy, University of Central Lancashire, Preston PR1 2HE, UK*

## ABSTRACT

Many circumbinary gas giant planets have been recently discovered. The formation mechanism of circumbinary planets on wide orbits is unclear. We investigate the formation of Delorme 1 (AB)b, a $13 \pm 5$ M$_J$ planet, orbiting its host binary at 84 au. The planet is accreting while having an estimated age of 40 Myr, which is unexpected, as this process should have ceased due to the dissipation of the protoplanetary disc. Using the smoothed particle hydrodynamics code SEREN, we model three formation scenarios for this planet. In Scenario I, the planet forms *in situ* on a wide orbit in a massive disc (by gravitational instability), in Scenario II closer to the binary in a massive disc (by gravitational instability), and in Scenario III much closer to the binary in a less massive disc (by core accretion). Planets in Scenario I stay at the observed separation and have mass accretion rates consistent with observed value, but their final mass is too high. In Scenario II, the planet reaches the observed separation through outward migration or scattering by the binary, and has mass accretion rate comparable to the observed; however, the planet mass is above the observed value. In Scenario III, the planet's final mass and mass accretion rate are comparable to the observed ones, but the planet's separation is smaller. We conclude that all models may explain some features of the observations but not all of them, raising questions about how gas is accreted on to the planet from its circumplanetary disc, and the presumed age of the system.

## 1 INTRODUCTION

Over 5600 exoplanets[1] have been confirmed since the discovery of 51 Pegasi b (Mayor & Queloz 1995). Some of these exoplanets are circumbinary, i.e. the planet orbits a binary star (also known as a P-type planet) (Dvorak 1984). The first circumbinary exoplanet discovered was Kepler-16b (Doyle et al. 2011), with over 40 others being documented since (NASA Exoplanet Archive 2024). The formation of these planets remains an open and interesting question.

The two widely accepted gas giant planet formation theories are (i) core accretion and (ii) gravitational instability. The core accretion model suggests that the planet core is formed through the accretion of pebbles and planetesimals within a gaseous disc (Goldreich & Ward 1973; Mizuno 1980; Bodenheimer & Pollack 1986; Pollack et al. 1996; Drążkowska et al. 2023). An outcome of this accretion is the attainment of a gaseous envelope if the core has a sufficient mass. This process has difficulty forming gas giants on wide orbits due to time this growth takes, $\sim 10$ Myr (Pollack et al. 1996), which is longer than the estimated lifetime of discs, 3–5 Myr (Wagner, Apai & Kratter 2019). The formation of gas giant planets is also possible through gravitational fragmentation of discs. A protoplanetary disc is gravitationally unstable when it satisfies the Toomre criterion

(Toomre 1964),

$$Q \equiv \frac{c_s \Omega}{\pi G \Sigma} \lesssim Q_{crit} \simeq 1–2, \tag{1}$$

where $Q$ is the Toomre parameter, $c_s$ is the sound speed, $\Omega$ is the angular frequency, $G$ is the gravitational constant, and $\Sigma$ is the surface density of the disc. An outcome of gravitational instability is fragmentation that has the potential to lead to the formation of gas giant planets, if the cooling time of the disc is sufficiently short, i.e. $\tau_c \lesssim 3\Omega^{-1}$ (Gammie 2001). These conditions can be satisfied at large disc radii where fragmentation is therefore likely (e.g. Boley 2009; Stamatellos & Whitworth 2009).

In this paper, we explore the origin of Delorme 1 (AB)b [also known as 2MASS J01033563-5515561 (AB)b], a circumbinary planet first observed by Delorme et al. (2013). This system comprises a 0.19 and a 0.17 M$_\odot$ binary, with components separated by 12 au. The planet orbiting this binary is a gas giant of mass $M_p = 13 \pm 5$ M$_\odot$, at 84 au (Delorme et al. 2013; Eriksson et al. 2020). Eriksson et al. (2020) reported the discovery of very strong H α, H β, and He I line emission, which suggests active accretion, despite the age of the system, $\sim 40$ Myr (Ringqvist et al. 2023). The accretion rate ranges from $3.4 \times 10^{-10}$ to $2.0 \times 10^{-8}$ M$_J$ yr$^{-1}$ (Betti et al. 2022). The unusually long accretion time-scale is at odds with the dispersal time of the disc. Betti et al. (2022) suggest the presence of a 'Peter Pan' disc, a long-lived protoplanetary disc, which may explain why the planet is still actively accreting.

The aim of this work is to investigate the formation of the wide-orbit, circumbinary giant planet, Delorme 1 (AB)b. We consider three possible scenarios for the formation of the planet. The first is an *in situ* formation, at the observed distance (∼85 au) in a massive disc. After the formation, the planet remains at this orbit without significant perturbation.

⋆ E-mail: MTeasdale1@uclan.ac.uk

[1]NASA Exoplanet Archive, doi:10.26133/NEA12 (accessed on 2024 May 21).

The second scenario is formation in a massive disc closer to the binary and then outward migration to its current orbit. The planet may migrate inwards or outwards depending on the torque exerted on it from the inner/outer disc. For Type I migration, the interaction between the planet and the disc does not significantly alter the structure of the disc, with this interaction leading to the movement of the planet inwards (Ward 1997; Tanaka, Takeuchi & Ward 2002; Teasdale & Stamatellos 2023). Type II migration occurs when the planet opens a gap in the disc (Ward 1997; Paardekooper et al. 2023). The migration of the planet occurs in an inward direction as the disc evolves, with the time-scale set by the disc's viscosity. However, a planet may migrate outwards due to the interaction between the planet and the gravitationally unstable outer edge of the gap within the disc (Lin & Papaloizou 2012; Cloutier & Lin 2013). Teasdale & Stamatellos (2023) show that a circumbinary gas giant planet in a massive disc has an initial phase of inward, Type I migration, followed by outward, non-standard Type II migration that may allow a planet to reach a wider separation from the binary.

For the third scenario, we examine formation in a less massive disc, inward migration, and a scattering event with the binary that sends the planet on to the wide orbit. Dynamical interactions with a close-in binary can alter the orbit of a circumbinary planet into a circumstellar one (Gong & Ji 2018). Higuchi & Ida (2017) stress the importance of the initial scattering location among other factors on the final location of a planet, based upon models of HD-131399 Ab. Matsumura, Brasser & Ida (2021) find that cold Jupiters (at ∼20 au) may have been scattered into eccentric orbits. Generally, this occurs by an interaction with another giant planet, although it is possible that such an event could be caused by a companion star. Veras & Tout (2012) find that planets residing at a few tens of au from a binary could escape from the system through scattering.

We will use smoothed particle hydrodynamics (SPH) simulations of planets embedded in discs to model the above three possible formation scenarios for Delorme 1 (AB)b. We describe the computational method in Section 2, and the simulation set-up in Section 3. In Section 4, we present the set of simulations performed, and in Section 5 we relate these with the observed properties of Delorme 1 (AB)b. We finally discuss the wider implications of this work and its conclusions in Section 6.

## 2 COMPUTATIONAL METHOD

We use the computational method described by Teasdale & Stamatellos (2023) to simulate the dynamics of the circumbinary planet Delorme 1 (AB)b. We use SEREN, an SPH code developed by Hubber et al. (2011). The simulations use an implementation of the radiative transfer method developed by Stamatellos et al. (2007).

As in Teasdale & Stamatellos (2023), the binary and giant planet are represented by sink particles having radii $R_{sink,\star} = 0.2$ au and $R_{sink,p} = 0.1$ au, respectively. The planet's sink radius is set to this value to ensure that it is smaller than its Hill radius.

## 3 SIMULATION SET-UP

We perform a set of 24 simulations of a giant planet embedded in a circumbinary disc (see Table 1). Our aim is to examine which of the three different formation scenarios mentioned in Section 1 better match the observed properties of Delorme 1 (AB)b (Delorme et al. 2013; Eriksson et al. 2020; Betti et al. 2022b,a; Ringqvist et al. 2023).

We assume a circumbinary disc that extends from $R_{in}^d = 10$ au to $R_{out}^d = 100$ au, which is represented by $5 \times 10^5$ SPH particles. We model two initial circumbinary disc masses, $M_D = 0.04 M_\odot$ (Scenarios I and II) and $M_D = 0.01 M_\odot$ (Scenario III). The higher disc mass is chosen so that the disc is close to being gravitationally unstable at $R > 30$ au (see Fig. 1). It is then expected that it will promote outward migration of the embedded planet (Stamatellos 2015; Stamatellos & Inutsuka 2018; Teasdale & Stamatellos 2023). The lower disc mass is gravitationally stable (see Fig. 1), so that no outward planet migration is expected. We assume the system is observed face-on. The binary components have masses of $M_1 = 0.19 M_\odot$ and $M_2 = 0.17 M_\odot$. We vary the binary eccentricity between $e_b = 0$ and $e_b = 0.6$, increasing the eccentricity by 0.2 each time. Finally, we use an initial separation of $\alpha_b = 10$ au and $a_b = 12$ au.

As in Teasdale & Stamatellos (2023), we set the initial surface density profile and disc temperature to

$$\Sigma_0(R) = \Sigma(1 \text{ au}) \left( \frac{R}{\text{au}} \right)^{-1} \tag{2}$$

and

$$T_0(R) = 250 \text{ K} \left( \frac{R}{\text{au}} \right)^{-0.5} + 10 \text{ K}, \tag{3}$$

where $\Sigma(1 \text{ au})$ is determined by the mass and radius of the disc, and $R$ is the distance from the centre of mass of the binary.

The disc is relaxed, i.e. evolved without the planet, for 3 kyr as per Teasdale & Stamatellos (2023) and subsequently the planet is embedded in it. The planet has initial mass $M_p = 1 M_J$ and a circular orbit.

## 4 THE MASS AND ORBITAL PROPERTIES OF THE CIRCUMBINARY PLANET

We will briefly discuss the evolution of the binary and the planet for each formation scenario. Fig. 2 shows the evolution of the surface density for a typical run (Run 5). The planet initially migrates inwards, with Type I migration, before opening a gap in the disc whereupon the direction of migration is reversed (i.e. non-standard Type II migration; Stamatellos 2015; Teasdale & Stamatellos 2023).

### 4.1 Scenario I: *in situ* formation in a massive disc

#### 4.1.1 Binary evolution

We find that the binary for this formation scenario maintains a separation consistent to the observed one (see Table 1). The binary eccentricity in all runs for this scenario increases due to interactions with the circumbinary disc. The binary mass ratio does not change, matching the observed mass ratio.

#### 4.1.2 Planet evolution

The planet is initially embedded in the disc at 85 au, after which it migrates rapidly inwards (see Fig. 3) before slowing and reversing direction, due to interaction with the gravitationally unstable gap edges (Stamatellos 2015; Stamatellos & Inutsuka 2018; Teasdale & Stamatellos 2023). In almost all cases, the planet is able to go back to its initial separation within the simulation runtime. Therefore, the planet simulated with this formation scenario is able to match the observed separation of Delorme 1 (AB)b (Delorme et al. 2013).

**Table 1.** The parameters of the 24 simulations performed. $M_D$ is the initial disc mass, $\alpha_p$ is the initial planetary semimajor axis, $e_b$ is the initial binary eccentricity, and $\alpha_b$ is the initial binary semimajor axis. $\alpha_b^f$ is the final binary semimajor axis, $q_b^f$ is the final binary mass ratio, and $e_b^f$ is the final binary eccentricity. $\alpha_p^f$ is the final planet semimajor axis, $M_p^f$ is the final planet mass, $e_p^f$ is the final planet eccentricity, $r_p^f$ is the range of the final planet separation $\left[ r_{min} = \alpha_p^f \left( 1 - e_p^f \right), r_{max} = \alpha_p^f \left( 1 + e_p^f \right) \right]$, and $\dot{M}_p^f$ is the final mass accretion rate on to the planet. (Note: *Final* refers to values at the end of the hydrodynamic simulation, i.e. at 20 kyr.)

| Scenario | Run | $M_D$ (M$_\odot$) | $\alpha_p$ (au) | $e_b$ | $\alpha_b$ (au) | $\alpha_b^f$ (au) | $q_b^f$ | $e_b^f$ | $\alpha_p^f$ (au) | $M_p^f$ (MJ) | $e_p^f$ | $r_p^f$ | $\dot{M}_p^f$ (MJ yr$^{-1}$) |
|---|---|---|---|---|---|---|---|---|---|---|---|---|---|
| I | 1 | 0.04 | 85 | 0 | 10 | 9.2 | 0.90 | 0.26 | 64 | 14 | 0.04 | 61–67 | $2.7 \times 10^{-4}$ |
| I | 2 | 0.04 | 85 | 0.2 | 10 | 9.5 | 0.90 | 0.32 | 74 | 17 | 0.12 | 65–83 | $3.9 \times 10^{-4}$ |
| I | 3 | 0.04 | 85 | 0.4 | 10 | 9.7 | 0.89 | 0.41 | 74 | 19 | 0.08 | 68–80 | $3.8 \times 10^{-4}$ |
| I | 4 | 0.04 | 85 | 0.6 | 10 | 9.5 | 0.90 | 0.55 | 62 | 20 | 0.07 | 58–66 | $2.5 \times 10^{-4}$ |
| I | 5 | 0.04 | 85 | 0 | 12 | 11.2 | 0.90 | 0.25 | 77 | 17 | 0.11 | 69–85 | $4.5 \times 10^{-4}$ |
| I | 6 | 0.04 | 85 | 0.2 | 12 | 11.6 | 0.90 | 0.32 | 76 | 16 | 0.02 | 74–78 | $3.2 \times 10^{-4}$ |
| I | 7 | 0.04 | 85 | 0.4 | 12 | 11.7 | 0.90 | 0.41 | 82 | 18 | 0.06 | 77–87 | $2.9 \times 10^{-4}$ |
| I | 8 | 0.04 | 85 | 0.6 | 12 | 11.5 | 0.90 | 0.53 | 79 | 18 | 0.05 | 75–83 | $2.4 \times 10^{-4}$ |
| II | 9 | 0.04 | 60 | 0 | 10 | 9.2 | 0.90 | 0.26 | 66 | 17 | 0.13 | 57–75 | $4.8 \times 10^{-4}$ |
| II | 10 | 0.04 | 60 | 0.2 | 10 | 9.6 | 0.90 | 0.31 | 74 | 18 | 0.10 | 67–81 | $3.8 \times 10^{-4}$ |
| II | 11 | 0.04 | 60 | 0.4 | 10 | 9.5 | 0.90 | 0.41 | 60 | 18 | 0.02 | 59–61 | $3.2 \times 10^{-4}$ |
| II | 12 | 0.04 | 60 | 0.6 | 10 | 9.6 | 0.91 | 0.54 | 58 | 18 | 0.04 | 56–60 | $2.2 \times 10^{-4}$ |
| II | 13 | 0.04 | 60 | 0 | 12 | 11.1 | 0.90 | 0.26 | 62 | 16 | 0.03 | 60–64 | $3.0 \times 10^{-4}$ |
| II | 14 | 0.04 | 60 | 0.2 | 12 | 12.3 | 0.90 | 0.26 | 49 | 18 | 0.14 | 42–56 | $5.8 \times 10^{-4}$ |
| II | 15 | 0.04 | 60 | 0.4 | 12 | 12.3 | 0.89 | 0.39 | 40 | 13 | 0.10 | 36–44 | $8.9 \times 10^{-5}$ |
| II | 16 | 0.04 | 60 | 0.6 | 12 | 11.6 | 0.91 | 0.54 | 85 | 25 | 0.35 | 55–115 | $2.8 \times 10^{-4}$ |
| III | 17 | 0.01 | 30 | 0 | 10 | 10.0 | 0.90 | 0.00 | 26 | 3 | 0.07 | 24–28 | $2.4 \times 10^{-5}$ |
| III | 18 | 0.01 | 30 | 0.2 | 10 | 10.0 | 0.90 | 0.21 | 30 | 3 | 0.01 | 30 | $2.3 \times 10^{-5}$ |
| III | 19 | 0.01 | 30 | 0.4 | 10 | 10.0 | 0.90 | 0.40 | 34 | 3 | 0.07 | 32–36 | $3.3 \times 10^{-5}$ |
| III | 20 | 0.01 | 30 | 0.6 | 10 | 10.0 | 0.89 | 0.60 | 56 | 6 | 0.14 | 48–64 | $8.6 \times 10^{-5}$ |
| III | 21 | 0.01 | 30 | 0 | 12 | 12.0 | 0.90 | 0.01 | 30 | 3 | 0.03 | 29–31 | $2.6 \times 10^{-5}$ |
| III | 22 | 0.01 | 30 | 0.2 | 12 | 12.1 | 0.90 | 0.21 | 38 | 4 | 0.05 | 36–40 | $1.9 \times 10^{-5}$ |
| III | 23 | 0.01 | 30 | 0.4 | 12 | 11.9 | 0.90 | 0.41 | 54 | 6 | 0.02 | 53–55 | $1.4 \times 10^{-4}$ |
| III | 24 | 0.01 | 30 | 0.6 | 12 | 12.0 | 0.90 | 0.59 | 31 | 1 | – | – | – |

The mass of the planet is consistent with the observed mass of Delorme 1 (AB)b (Ringqvist et al. 2023), apart from two simulations (see Fig. 3). However, as the planet is still accreting material from the disc, it may surpass this mass (see Section 5).

## 4.2 Scenario II: formation closer to the binary in a massive disc

### 4.2.1 Binary evolution

The binary sees no significant change in separation during the simulation runtime (see Table 1). While different runs do show a decrease in separation, this change is small, with the eccentricity of the binary still making the separation consistent with the observed separation. Similarly, the binary mass ratio does not change significantly throughout the simulations.

### 4.2.2 Planet evolution

The evolution of the planet's orbit shows two different patterns for the two binary separations that we examine here (see Fig. 4). For the runs with binary separation of 10 au (Runs 9–12), the planet initially migrates inwards before slowing down and reversing direction, as in Scenario I. The reversal in the migration direction is due to the interaction between the planet and the gravitationally unstable disc just inside and outside of the planet-induced gap (Teasdale & Stamatellos 2023). For the runs with binary separation of 12 au (Runs 13–16), the planet migrates inwards, enters the cavity around the binary, and gets scattered by the binary on to a wide orbit (apart from Run 13 that follows the previous pattern). This is because the binary-induced cavity is wider in the case of the 12 au binary than for the 10 au binary; Lubow & Artymowicz (1996) estimate the size of the cavity to be ∼2–3 times the binary separation.

The most notable run is that of $\alpha_b = 10$ au and $e_b = 0.2$ (Run 10), in which the planet undergoes inward migration for ∼2.5 kyr before slowing down and reversing migration (going from Type I to a non-standard Type II). As a result of this non-standard Type II migration, the planet is able to reach a final semimajor axis of 74 au. We find the final separation to be 81 au (see Table 1), which is comparable to the observed separation (Delorme et al. 2013). This run is also notable as the final mass of the planet is consistent with the observed value. However, we do expect that this mass will continue to increase due to ongoing accretion on to the planet from the disc.

The simulation with parameters $\alpha_b = 10$ au and $e_b = 0$ (Run 9) follows a similar pattern to the run discussed previously, i.e. rapid inward migration followed by a slow outward migration. As with the previous discussed simulation, the planet here also has a mass similar to the observed value at the end of the runtime.

The best match to the observed separation is the run with $\alpha_b = 12$ au and $e_b = 0.6$ (Run 16). In this run, the planet reaches a final semimajor axis of $\alpha_p = 85$ au, a value very close to the observed separation of 84 au (Delorme et al. 2013). The planet reaches this orbit not through outward migration but instead through outward scattering by the binary. The scattering event occurs at ∼7.5 kyr, seen in Fig. 4. As a result of the scattering event, the planet's eccentricity is also affected (see Fig. 4). Due to the scattering into the outer regions of the disc, the planet's mass increases substantially. A similar scattering event occurs in the simulation with parameters $\alpha_b = 12$ au and $e_b = 0.2$ (Run 14). However, the planet is unable to reach a similar wide orbit.

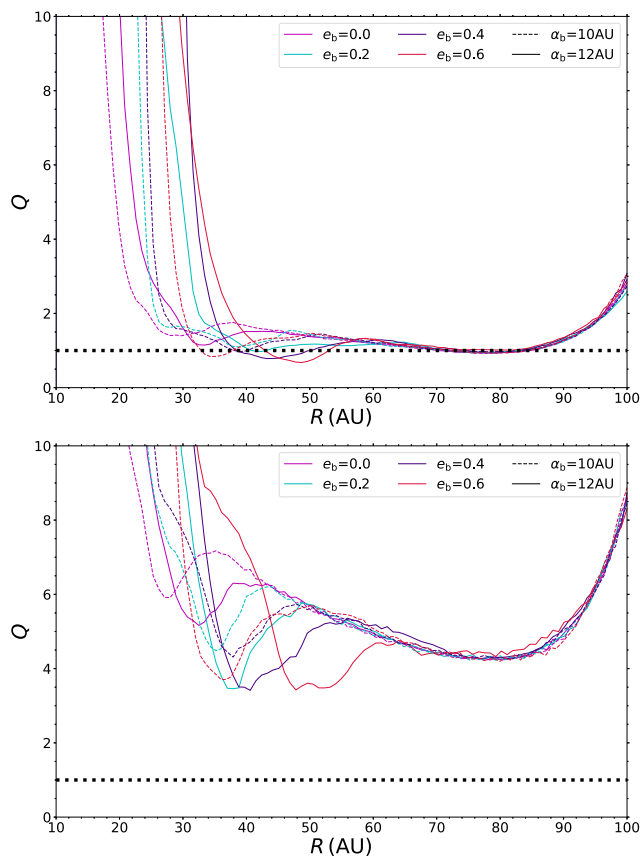

**Figure 1.** The initial Toomre parameter for the disc in the Scenario I/II ($M_D = 0.04\,M_\odot$; top) and Scenario III ($M_D = 0.01\,M_\odot$; bottom) simulations plotted against the distance from the centre of mass of the binary. The disc in Scenario I/II is gravitationally unstable outside $\sim$30 au. In contrast, the disc for Scenario III is gravitationally stable ($Q \gtrsim 4$).

We note that despite being able to replicate the observed planet separation, we were unable to find a combination of parameters that allowed the planet to attain a mass comparable to the one observed. As the planet is still accreting material from the disc at the end of the simulation, the planet's mass is expected to grow beyond the observed value (see Section 5).

### 4.3 Scenario III: formation close to the binary in a low-mass disc

#### 4.3.1 Binary evolution

We find all runs with this formation scenario to be in agreement with the observed binary parameters. There is little to no change in the separation, mass ratio, or eccentricity over the simulation runtime (see Table 1).

#### 4.3.2 Planet evolution

Embedding the planet on a close orbit within a much less massive disc, to replicate a possible formation by the core-accretion model, yields a different evolutionary path to Scenario I and Scenario II. Outward migration through an interaction with a gravitationally unstable disc is impossible. However, as the planet is much closer to the binary, scattering becomes a much more likely outcome. Indeed, we see this happening in four runs (see Fig. 5).

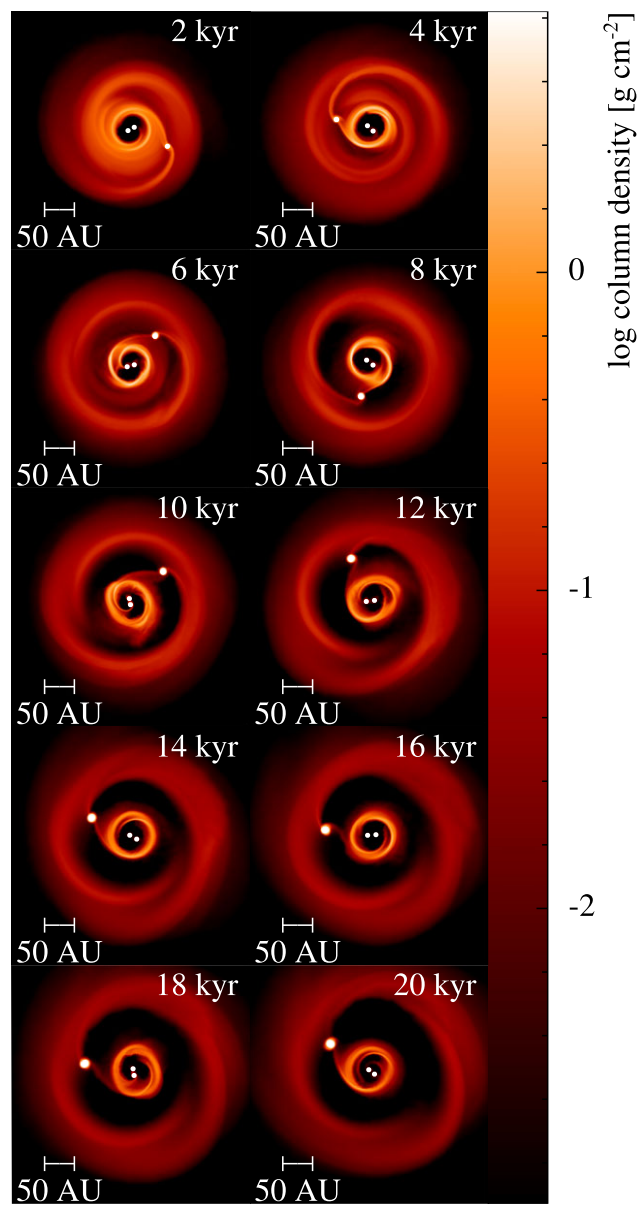

**Figure 2.** The evolution of the disc surface density ($g\,cm^{-2}$) for Run 5 (Scenario I) listed in Table 1. A 1 $M_J$ planet is embedded at 85 au in a 0.04 $M_\odot$ disc, around a binary with separation $\alpha_b = 12$ au and eccentricity $e_b = 0$. The disc–planet interaction is shown from 2 kyr until the end of the simulation at 20 kyr.

For example, in the simulation with parameters $\alpha_b = 10$ au and $e_b = 0.6$ (Run 20; see Table 1), the planet is scattered shortly after it is embedded in the disc and reaches a separation of $\sim$120 au before moving closer to the binary again. A stable orbit is achieved within 5 kyr of reaching its apoapsis. By the end of the simulation, the planet mass is below the observed value (see Fig. 5).

Another similarly scattered planet is that of the run with $\alpha_b = 12$ au and $e_b = 0.4$ (Run 23). Despite not reaching the same semimajor axis as in the previously discussed simulation, this planet is perturbed shortly after it is embedded and reaches an orbit wider than its initial. The outward motion leads to an apoapsis at $\sim$90 au before the planet settling into an orbit at 54 au. We note that the planet discussed here and the one discussed previously reach an almost identical final stable orbit despite a significant difference in evolutionary path and apoapsis.

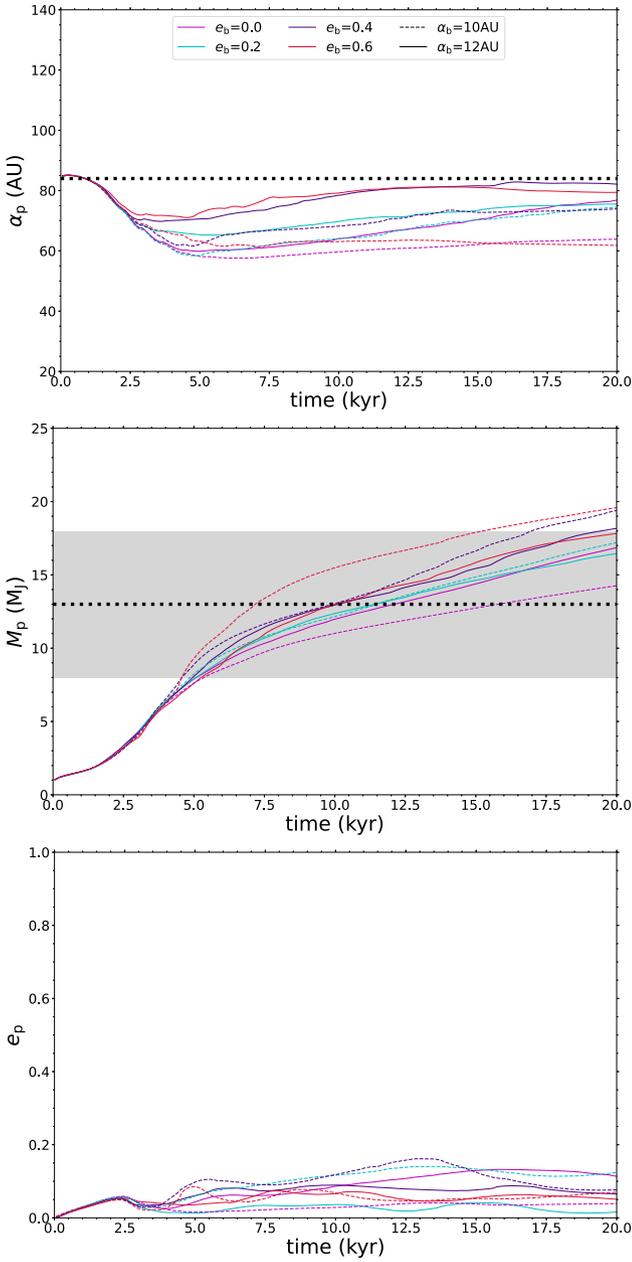

**Figure 3.** The semimajor axis (top), the mass (middle), and the eccentricity (bottom) of the planet for Scenario I (i.e. $M_D = 0.04\,M_\odot$ and $\alpha_p = 85$ au) plotted against the time. The dashed line on the top graph denotes the observed separation (Delorme et al. 2013). The dashed line on the middle graph indicates the planet mass estimated by Ringqvist et al. (2023), with the light grey area denoting the error of $\pm 5\,M_J$ on this value.

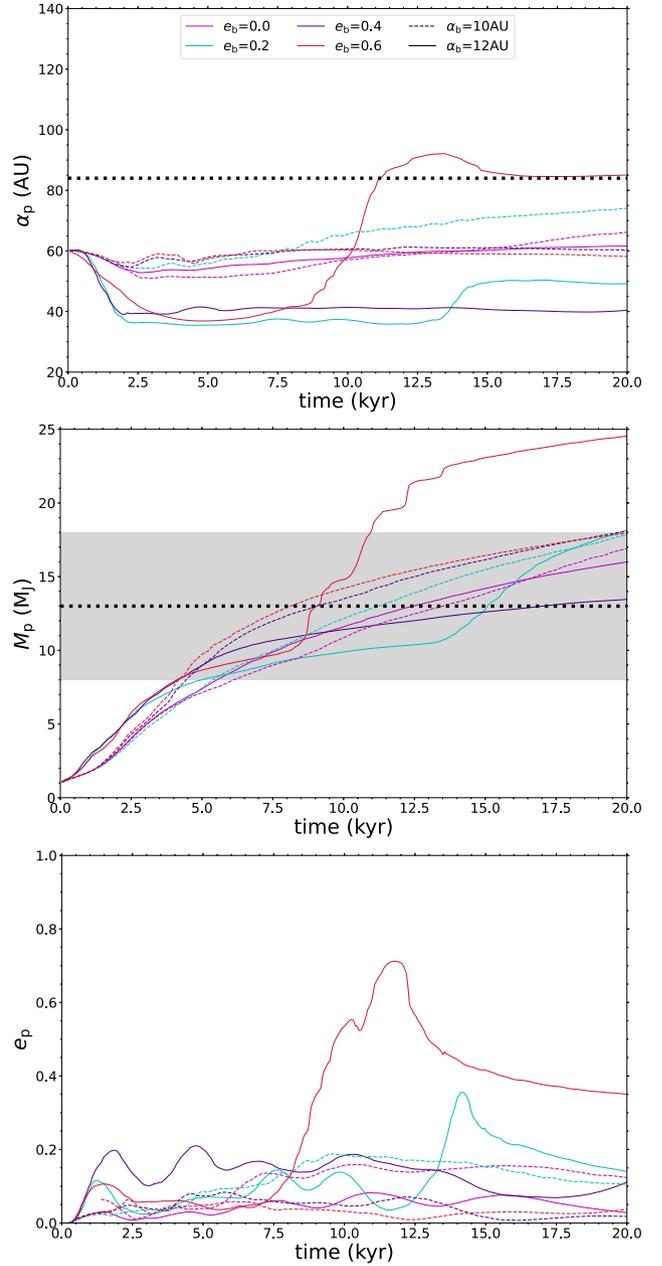

**Figure 4.** The semimajor axis (top), the mass (middle), and the eccentricity (bottom) of the planet for Scenario II (i.e. $M_D = 0.04\,M_\odot$ and $\alpha_p = 60$ au) plotted against the time. The dashed line on the top graph denotes the observed separation (Delorme et al. 2013). The dashed line on the middle graph indicates the planet mass estimated by Ringqvist et al. (2023), with the light grey area denoting the error of $\pm 5\,M_J$ on this value.

In the simulations with parameters $\alpha_b = 12$ au and $e_b = 0.2$ (Run 22; see Table 1), the planet does not reach a separation close to the observed value, but we note that scattering events did take place. Due to the stochastic nature of the scattering interaction, the final orbit of the planet may vary significantly in this scenario.

Finally, we note the simulation with parameters $\alpha_b = 12$ au and $e_b = 0.6$ (Run 24; see Table 1). After we embed the planet in the disc, it is dynamically scattered by the binary and ejected from the system at $\sim 0.5$ kyr (see Fig. 5 and Table 1).

## 5 THE ACCRETION RATE ON TO THE CIRCUMBINARY PLANET

One of the most interesting features of the Delorme 1 (AB)b circumbinary planet is that it shows signs of accretion ($3.4 \times 10^{-10}$ to $2.0 \times 10^{-8}\,M_J\,\mathrm{yr}^{-1}$; Eriksson et al. 2020; Betti et al. 2022; Ringqvist et al. 2023). The accretion rate on to the planet at 20 kyr (end time of the SPH simulations) is shown in Fig. 6, and summarized in Table 2. The accretion rates for Scenarios I and II (more massive discs; $M_D = 0.04 \, M_\odot$) are of the order of $10^{-4} \, M_J \, yr^{-1}$, whereas for Scenario III (less massive discs; $M_D = 0.01 \, M_\odot$) of the order of $10^{-5} \, M_J \, yr^{-1}$, i.e. an order of magnitude lower, with the exception of Runs 20 and 23, in which the planet is scattered on a wide orbit; in these two cases, the accretion rate is of the order of $10^{-4} \, M_J \, yr^{-1}$, but still 2–3 times lower than the accretion rate for Scenarios I and II.

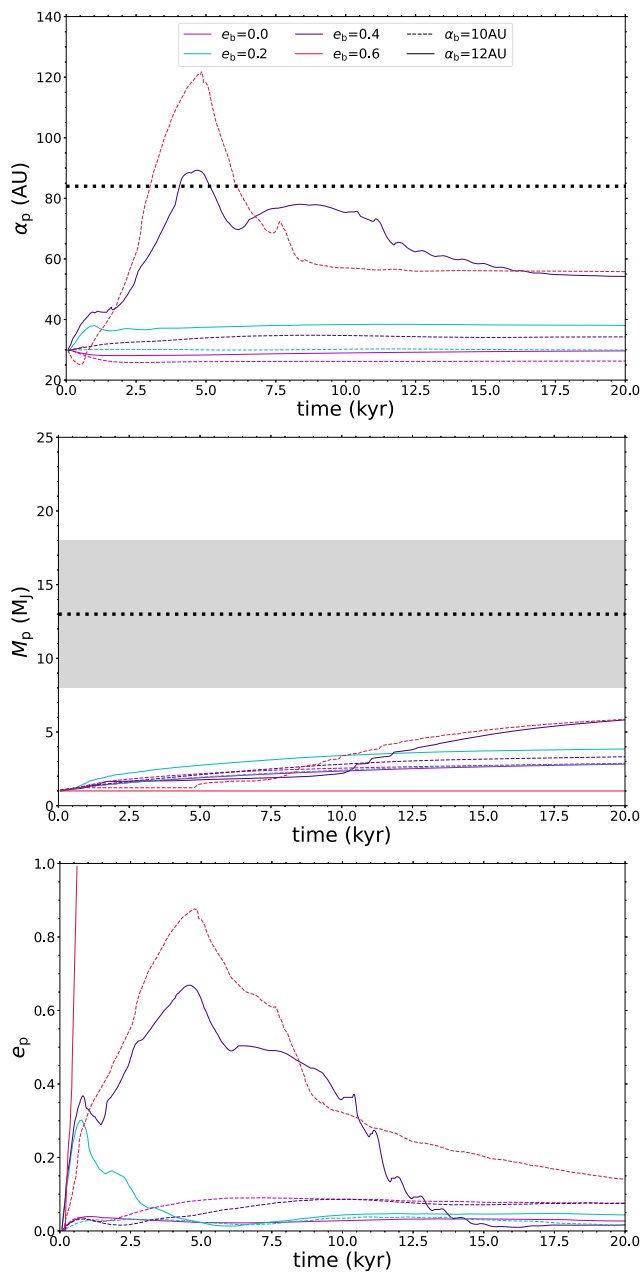

**Figure 5.** The semimajor axis (top), the mass (middle), and the eccentricity (bottom) of the planet for Scenario III (i.e. $M_D = 0.01 \, M_\odot$ and $\alpha_p = 30$ au) plotted against the time. The dashed line on the top graph denotes the observed separation (Delorme et al. 2013). The dashed line on the middle graph indicates the planet mass estimated by Ringqvist et al. (2023), with the light grey area denoting the error of $\pm 5 \, M_J$ on this value.

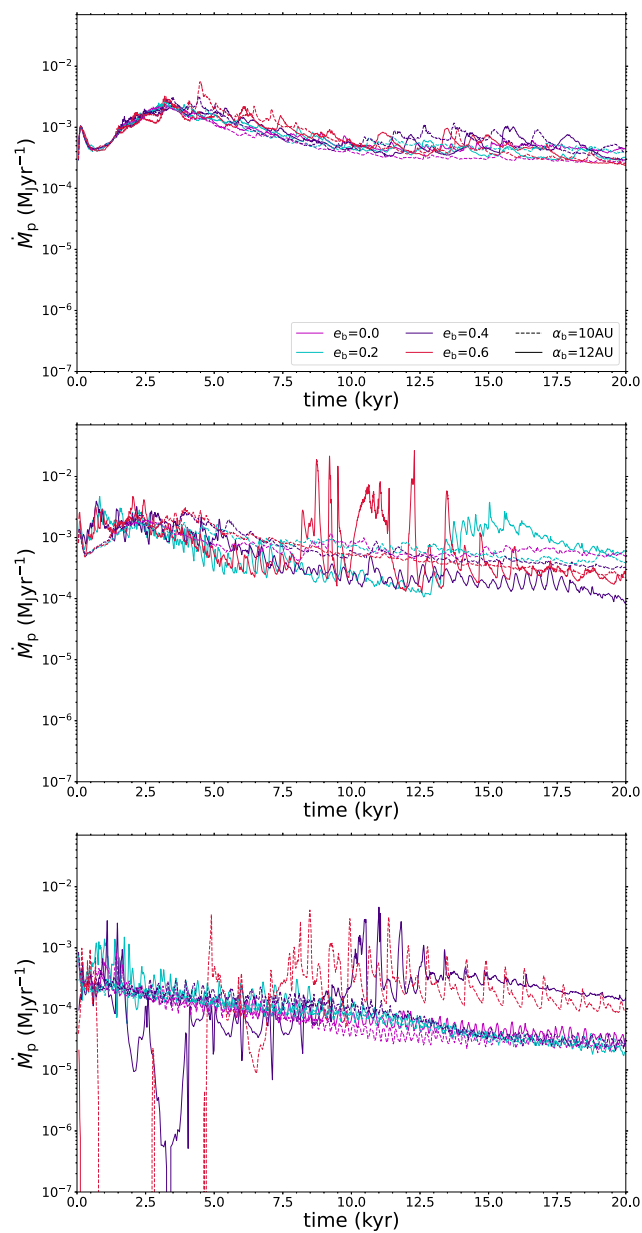

**Figure 6.** The accretion rate on to the planet for Scenario I (top), Scenario II (middle), and Scenario III (bottom) plotted against time.

Before discussing the details of the accretion rates on to the planet for the different formation scenarios examined here, we need to estimate the mass accretion rate on to the planet at the presumed age of the system (40 Myr), in order to compare with observations. We assume that the planet accretes gas from its circumplanetary disc and we then calculate the evolution of the accretion rate on to the planet assuming that the circumplanetary disc evolves viscously. We further assume that the circumplanetary viscosity is independent of time and can be expressed as a power law in $R$, $\nu \propto R^\gamma$ (Lynden-Bell & Pringle 1974; Hartmann et al. 1998; Stamatellos & Herczeg 2015). Then, the circumplanetary viscous evolution time-scale is

$$ t_v = 8 \times 10^4 \left( \frac{\alpha}{10^{-2}} \right)^{-1} \left( \frac{R_0}{10 \, au} \right) \left[ \frac{M_p(t_0)}{524 \, M_J} \right]^{\frac{1}{2}} \left( \frac{T_d}{10 \, K} \right)^{-1} yr, \quad (4) $$

where $\alpha$ is the viscosity parameter of the circumplanetary disc, $R_0$ is radius of the circumplanetary disc in which 60 per cent of the mass is contained, $M_p(t_0)$ is the mass of the planet, and $T_d$ is the

**Table 2.** The long-term evolution of the planet parameters for the 24 simulations performed. $\dot{M}_p^t$ is the mass accretion rate on to the planet at 20 kyr, $\dot{M}_p^{40\,Myr}$ is the mass accretion rate on to the planet at 40 Myr, $M_p^t$ is the planet mass at 20 kyr, and $M_p^{40\,Myr}$ is the planet mass at 40 Myr. $\alpha$ is the assumed viscosity parameter of the circumplanetary disc (see equation 5).

| Scenario | Run | $\dot{M}_p^t$ (M$_J$ yr$^{-1}$) | $\dot{M}_p^{40\,Myr}$ (M$_J$ yr$^{-1}$) | | | | $M_p^t$ | $M_p^{40\,Myr}$ | | | |
|---|---|---|---|---|---|---|---|---|---|---|---|
| | | | $\alpha=10^{-1}$ | $\alpha=10^{-2}$ | $\alpha=10^{-3}$ | $\alpha=10^{-4}$ | | $\alpha=10^{-1}$ | $\alpha=10^{-2}$ | $\alpha=10^{-3}$ | $\alpha=10^{-4}$ |
| I | 1 | $2.7\times10^{-4}$ | $3.1\times10^{-9}$ | $4.2\times10^{-9}$ | $1.9\times10^{-8}$ | $3.8\times10^{-7}$ | 14 | 25 | 27 | 50 | 256 |
| I | 2 | $3.9\times10^{-4}$ | $4.5\times10^{-9}$ | $6.6\times10^{-9}$ | $3.9\times10^{-8}$ | $8.5\times10^{-7}$ | 17 | 33 | 37 | 81 | 478 |
| I | 3 | $3.8\times10^{-4}$ | $4.5\times10^{-9}$ | $6.8\times10^{-9}$ | $4.3\times10^{-8}$ | $9.6\times10^{-7}$ | 19 | 35 | 40 | 87 | 518 |
| I | 4 | $2.5\times10^{-4}$ | $2.9\times10^{-9}$ | $4.2\times10^{-9}$ | $2.3\times10^{-8}$ | $5.0\times10^{-7}$ | 20 | 30 | 32 | 59 | 301 |
| I | 5 | $2.8\times10^{-4}$ | $2.8\times10^{-9}$ | $7.7\times10^{-9}$ | $4.6\times10^{-8}$ | $1.0\times10^{-6}$ | 17 | 27 | 30 | 60 | 327 |
| I | 6 | $3.2\times10^{-4}$ | $3.8\times10^{-9}$ | $5.5\times10^{-9}$ | $3.2\times10^{-8}$ | $6.9\times10^{-7}$ | 16 | 30 | 33 | 69 | 397 |
| I | 7 | $2.9\times10^{-4}$ | $3.5\times10^{-9}$ | $5.3\times10^{-9}$ | $3.5\times10^{-8}$ | $7.9\times10^{-7}$ | 18 | 30 | 34 | 72 | 417 |
| I | 8 | $2.4\times10^{-4}$ | $2.8\times10^{-9}$ | $4.2\times10^{-9}$ | $2.7\times10^{-8}$ | $6.0\times10^{-7}$ | 18 | 27 | 30 | 60 | 327 |
| II | 9 | $4.8\times10^{-4}$ | $5.6\times10^{-9}$ | $7.9\times10^{-9}$ | $4.2\times10^{-8}$ | $8.7\times10^{-7}$ | 17 | 36 | 41 | 89 | 527 |
| II | 10 | $3.8\times10^{-4}$ | $4.5\times10^{-9}$ | $6.6\times10^{-9}$ | $4.0\times10^{-8}$ | $8.7\times10^{-7}$ | 18 | 33 | 38 | 82 | 485 |
| II | 11 | $3.2\times10^{-4}$ | $3.7\times10^{-9}$ | $5.2\times10^{-9}$ | $2.7\times10^{-8}$ | $5.5\times10^{-7}$ | 18 | 31 | 34 | 65 | 346 |
| II | 12 | $2.2\times10^{-4}$ | $2.6\times10^{-9}$ | $3.5\times10^{-9}$ | $1.8\times10^{-8}$ | $3.6\times10^{-7}$ | 18 | 27 | 29 | 49 | 237 |
| II | 13 | $3.0\times10^{-4}$ | $3.4\times10^{-9}$ | $4.7\times10^{-9}$ | $2.3\times10^{-8}$ | $4.6\times10^{-7}$ | 16 | 28 | 31 | 57 | 299 |
| II | 14 | $5.8\times10^{-4}$ | $6.7\times10^{-9}$ | $8.8\times10^{-9}$ | $3.9\times10^{-8}$ | $7.5\times10^{-7}$ | 18 | 41 | 46 | 91 | 514 |
| II | 15 | $8.9\times10^{-5}$ | $1.0\times10^{-9}$ | $1.2\times10^{-9}$ | $3.9\times10^{-9}$ | $6.1\times10^{-8}$ | 13 | 17 | 17 | 22 | 64 |
| II | 16 | $2.8\times10^{-4}$ | $3.4\times10^{-9}$ | $5.8\times10^{-9}$ | $4.8\times10^{-8}$ | $1.1\times10^{-6}$ | 25 | 36 | 41 | 90 | 519 |
| III | 17 | $2.4\times10^{-5}$ | $2.7\times10^{-10}$ | $2.8\times10^{-10}$ | $3.9\times10^{-10}$ | $1.9\times10^{-9}$ | 3 | 4 | 4 | 4 | 6 |
| III | 18 | $2.3\times10^{-5}$ | $2.6\times10^{-10}$ | $2.7\times10^{-10}$ | $3.8\times10^{-10}$ | $2.1\times10^{-9}$ | 3 | 4 | 4 | 4 | 6 |
| III | 19 | $3.3\times10^{-5}$ | $3.7\times10^{-10}$ | $3.9\times10^{-10}$ | $6.0\times10^{-10}$ | $4.0\times10^{-9}$ | 3 | 5 | 5 | 5 | 9 |
| III | 20 | $8.6\times10^{-5}$ | $9.8\times10^{-10}$ | $1.1\times10^{-9}$ | $2.8\times10^{-9}$ | $3.6\times10^{-8}$ | 6 | 9 | 10 | 13 | 42 |
| III | 21 | $2.6\times10^{-5}$ | $2.9\times10^{-10}$ | $3.0\times10^{-10}$ | $4.3\times10^{-10}$ | $2.3\times10^{-9}$ | 3 | 4 | 4 | 4 | 7 |
| III | 22 | $1.9\times10^{-5}$ | $2.1\times10^{-10}$ | $2.3\times10^{-10}$ | $3.8\times10^{-10}$ | $3.0\times10^{-9}$ | 4 | 5 | 5 | 5 | 8 |
| III | 23 | $1.4\times10^{-4}$ | $1.6\times10^{-9}$ | $1.8\times10^{-9}$ | $4.4\times10^{-9}$ | $5.6\times10^{-8}$ | 6 | 11 | 12 | 17 | 62 |
| III | 24* | – | – | – | – | – | 1 | – | – | – | – |

*Note.* *The planet in this run is quickly (within 0.5 kyr) dynamically ejected from the system after interacting with the binary, and the accretion stops.

temperature of the disc at its outer edge. For simplicity, we set $R_0$ to be 1/4 $R_H$, where $R_H$ is the Hill radius of the planet at 20 kyr. This corresponds to 75 per cent of the size of the circumplanetary disc that is estimated to be 1/3 $R_H$ (Ayliffe & Bate 2009). Using the formulation of Hartmann et al. (1998) and assuming that $\gamma=1$, we can relate the accretion rate on to the planet at any time $t$ with the accretion rate on to the planet at $t=20$ kyr (at the end of the SPH simulations),

$$\dot{M}_p(t) = \dot{M}_p(t_0)\frac{\left(\frac{t}{t_\nu}+1\right)^{-\frac{3}{2}}}{\left(\frac{t_0}{t_\nu}+1\right)^{-\frac{3}{2}}}, \qquad (5)$$

where $\dot{M}_p(t_0)$ is the mass accretion rate on to the planet at 20 kyr. We also calculate the evolution of the planet mass by integrating the above equation,

$$M_p(t) = M_p(t_0) + 2\dot{M}_p(t_0)(t_0+t_\nu)\left[1 - \left(\frac{\frac{t}{t_\nu}+1}{\frac{t_0}{t_\nu}+1}\right)^{-\frac{1}{2}}\right], \qquad (6)$$

where $M_p(t_0)$ is the mass of the planet at 20 kyr. The estimated accretion rate and planet mass at the presumed age of the circumbinary system (40 Myr; Delorme et al. 2013; Ringqvist et al. 2023) for different values of the viscosity parameter $\alpha$ ($10^{-4}$, $10^{-3}$, $10^{-2}$, and $10^{-1}$) are shown in Table 2.

In the Scenario I simulations (see Fig. 6), we see a sharp increase of the mass accretion rate soon after the planet is embedded in the disc as the planet is clearing up a gap at its orbit and migrates inwards.

There is a subsequent decrease of the accretion rate as the gap has been opened up and then a slower increase to a peak at ∼3.5 kyr as the planet starts migrating outwards. When the planet orbit stabilizes, the circumplanetary disc is not being vigorously fed by the circumstellar disc and it slowly depletes on to the planet, with the accretion rate slowly dropping. In the runs where scattering occurs, the behaviour of the accretion rate is similar. The estimated accretion rate on to the planet at the observed age of the system depends on the assumed viscosity parameter of the circumplanetary disc. For lower $\alpha$, the accretion rates drop slower resulting in a higher accretion rate at 40 Myr, but at the same time this results in a higher planet mass. The models that are broadly more consistent with the observed planet mass and accretion rate are those with $\alpha=10^{-1}$ and $\alpha=10^{-2}$. These give accretion rates of $(3.1–4.2)\times10^{-9}$ M$_J$ yr$^{-1}$ (consistent with observations) and planet mass of 25–27 M$_J$ (higher than observed). Models with $\alpha=10^{-3}$ and $\alpha=10^{-4}$ give unrealistically large planet masses. This behaviour is mirrored in the simulations of Scenario II.

In Scenario III, the disc has lower mass, resulting in lower accretion rate on to the planet, with the majority of these runs showing accretion rates smaller by an order of magnitude than in Scenarios I and II, $(2.1–3.7)\times10^{-9}$ M$_J$ yr$^{-1}$ (see Fig. 6 and Table 2). However, in the runs in which the planet gets scattered by the binary and reaches a wide orbit, comparable to observations (Runs 20 and 23), the accretion rate on to the planet, $(0.98–1.6)\times10^{-9}$ M$_J$ yr$^{-1}$, is higher but still lower by a few times than that in the Scenario I and II runs. In these two runs, the accretion rate initially varies significantly; it starts off similar to the accretion rates of the other runs in this scenario, but as the planet gets scattered in the outer disc region the accretion rate drops considerably. After the planet returns within the disc on a stable orbit, its accretion rate increases again. The final planet mass in these two runs, 9–17 $M_J$, is consistent with observations, for models with $\alpha = 10^{-3}$, $10^{-2}$, and $10^{-1}$. A low alpha ($\alpha = 10^{-3}$) is also favoured by Betti et al. (2022), when comparing their accretion rate observations to the models of Stamatellos & Herczeg (2015).

# 6 CONCLUSIONS

We used the SPH code SEREN to investigate the potential origin of Delorme 1 (AB)b. We presented three formation scenarios for this object: (I) an *in situ* formation in a massive disc ($M_D = 0.04\,M_\odot$); (II) a closer in formation than Scenario I and outward migration in a massive disc ($M_D = 0.04\,M_\odot$); and (III) formation closer to the binary in a lower mass disc ($M_D = 0.01\,M_\odot$). The first two scenarios relate to marginally unstable discs ($Q_{min} \sim 1$), whereas the third scenario relates to stable discs ($Q_{min} \sim 4$). Therefore, Scenarios I and II may be thought to represent formation by gravitational instability, whereas Scenario III to represent formation by core accretion. We note, however, that we do not study the formation of the planet, but only its evolution after it has been formed. The initial planet mass was set to 1 $M_J$. We then calculated the evolution of the mass, orbital radius, and accretion rate on to the planet for these different scenarios and for varying separations and eccentricities of the binary.

In Scenario I, the planet shows an initial phase of inward migration before starting migrating outwards, close to its initial orbital radius. The planet is able to match the observed separation in the majority of the runs. The planet mass at the end of the simulation (20 kyr) is 14–20 $M_J$, i.e. near the upper limit of the observed value, and the accretion rate on to it is (2.4–4.5) × $10^{-4}$ $M_J$ yr$^{-1}$. In Scenario II, the results are similar; the planet initially migrates inwards, opens up a gap, and then migrates outwards matching the observed orbital radius, having mass near the upper limit of the observed mass of 17–18 $M_J$, and accretion rate of (1.8–4.8) × $10^{-4}$ $M_J$ yr$^{-1}$. Outward dynamical scattering is also possible in this case if during the planet's inward migration it reaches near ~3 times the separation of the binary. In Scenario III, there are two paths. In most runs, the planet remains close to its initial separation, i.e. ~30 au from the binary, 26–38 au, with its final mass below the observed value, 3–4 $M_J$. In three of the runs, the planet gets either dynamically scattered from a binary to a wider orbit (two runs) or ejected from the system (one run). The orbital radius of the planet in the runs where scattering happens is ~55 au, which is below the observed value. However, due to the stochastic nature of the scattering, a wider orbit closer to the observed one may also be possible. The planet mass in these two runs (6 $M_J$) is just below the observed lower limit, whereas the accretion rate is just a few times lower than that in Scenarios I and II, (1.8–4.8) × $10^{-4}$ $M_J$ yr$^{-1}$.

To facilitate a better comparison with observations, we used a simple viscous disc model to determine the projected planet mass and accretion rate at the estimated age of the system (~40 Myr), assuming that the planet accretes from its circumplanetary disc. For the Scenario I and II simulations (in which the planet exhibits high accretion rates at the end of the SPH simulation), we are able to match the observed accretion rates for viscosity parameters of $10^{-2}$ and $10^{-1}$, estimating ~(3–4) × $10^{-4}$ $M_J$ yr$^{-1}$; however, the calculated mass is above the observed one by at least 7 $M_J$. For the two runs in Scenario III in which the planet ends up on a wide orbit, we find accretion rate and planet mass compatible with observations for models with viscosity parameters of $10^{-3}$, $10^{-2}$, and $10^{-1}$. We note that there is a high uncertainty in the observed accretion rate on to the planet, ranging from 3.4 × $10^{-10}$ to 2.0 × $10^{-8}$ $M_J$ yr$^{-1}$ (Eriksson et al. 2020; Betti et al. 2022; Ringqvist et al. 2023). This may be a consequence of the variability in the episodic accretion rate over short time-scales. Variability at similar magnitudes and over similar time-scales has been reported for other very low mass accretors (e.g. Demars et al. 2023). In the best-match cases that we highlighted above, we get accretion rates of ~(1–4) × $10^{-9}$ $M_J$ yr$^{-1}$; therefore, our models do not support accretion rates as high as 2.0 × $10^{-8}$ $M_J$ yr$^{-1}$, if the age of the system is 40 Myr.

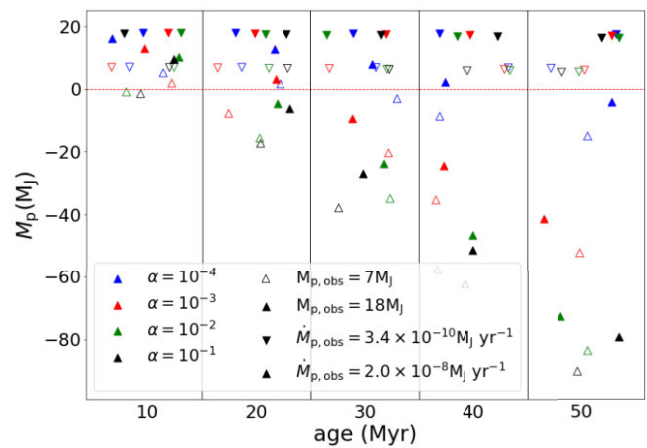

**Figure 7.** The initial planet mass $M_p$ needed in order to achieve the planet's final mass, after accreting gas over the age of the system. We use two fixed accretion rates (the minimum and maximum estimates from observations) that increase going backwards in time. We also use two values for the observed planet mass (minimum and maximum estimates), five values for the age of the system (10–50 Myr), and four different values of the $\alpha$ viscosity parameter (as marked on the graph). Colours correspond to different values of viscosity parameter $\alpha$, filled/unfilled symbols to different observed planet masses, and different symbols to different observed accretion rates. Negative planet masses correspond to forbidden combinations of parameters. (Note that all values within each age column correspond to the same age, but they have been spread horizontally across the column for better visibility.)

Considering the uncertainties in the estimated mass of the planet, the accretion rate on to it, the age of the system (Eriksson et al. 2020), and our incomplete understanding of how accretion of gas happens on to the planet from its circumplanetary disc (as this is described by the $\alpha$ parameter in the viscous evolution model), we perform a simple analysis to investigate what combination of values may provide a consistent description of the properties of this planet. We use the minimum and maximum observed values for the planet mass and accretion rate, and using the viscous evolution model described in Section 5 we go backwards in time to find the initial mass of the planet (just after its formation), for five different assumed ages of the system (from 10 to 50 Myr) and for four different values of the viscosity parameter $\alpha$. Effectively, we assume an accretion rate that increases when going back in time and calculate what the initial mass of the planet needs to be, so that the added mass due to accretion gives the observed value of the planet mass.

The results of this analysis are shown in Fig. 7. Negative values for the planet initial mass mean that too much mass is accreted over the age of the system, and therefore the corresponding combination of parameters is not possible. We see from Fig. 7 that high observed accretion rates are not compatible with a system age of 40 Myr, apart (marginally) from the case of $\alpha = 10^{-4}$ and a current planet mass of 18 $M_J$. Generally speaking, a high planet mass is compatible with a wider range of parameters. If the system is younger, then higher accretion rates and lower current planet masses are possible.

Therefore, if the high accretion rates reported are indeed accurate (Ringqvist et al. 2023), then the system may be younger than implied from its membership in Tucana–Horologium cluster, or it may even not belong to this cluster.

The inherent assumption of the viscous disc model that we used is that accretion on to the planet happens through its circumplanetary disc that behaves as a traditional accretion disc (e.g. Pringle 1981). However, simulations have shown the existence of complex flows within circumplanetary discs as they are fed with gas from the circumstellar (or circumbinary in our case) disc (Tanigawa, Ohtsuki & Machida 2012; Gressel et al. 2013). It is also contested whether the value of the viscosity parameter provided by the magneto-rotational instability in circumplanetary discs is high ($\alpha = 10^{-2}$; Gressel et al. 2013) or low (Fujii et al. 2014; Szulágyi et al. 2014). Therefore, more detailed models of the gas accretion on to the planet are needed for safer estimates.

Alternative processes that have not been considered here may also be possible. For example, migration of the planet to its current location may be achieved through planet–planet scattering. This scenario would require two giant planets forming in the system and undergoing planet–planet interactions. Such interactions could lead to the ejection of one planet from the system while the other gets scattered into the outer disc (Gong 2017).

We conclude that the three models examined here may explain specific features of the observations of Delorme 1 (AB)b, but not all at the same time. Therefore, we cannot exclude any of the presented formation scenarios, although our models show that higher planet accretion rate is more compatible with formation in a higher mass disc, possibly by gravitational fragmentation (see also Stamatellos & Herczeg 2015). Moreover, although dynamical scattering by the binary may reproduce the observed orbital separation of the planet, there is a stochastic element in this process making it rather rare, whereas formation by gravitational instability consistently produces planets at such wide orbital radii. Better constraints of the observed properties of the system are needed in order to pin down the formation mechanism of this planet.

## ACKNOWLEDGEMENTS

DS acknowledges support from STFC grant ST/Y002741/1. The simulations were performed using the UCLan High Performance Computing (HPC) facility. We thank David Hubber for the development of SEREN. Surface density plots were produced using SPLASH (Price 2007).

## DATA AVAILABILITY

The simulation data used for this paper can be provided by contacting the authors.

This paper has been typeset from a TeX/LaTeX file prepared by the author.